# A Formalization of Polytime Functions[*]

Sylvain Heraud[1] and David Nowak[2,**]

[1] INRIA Sophia Antipolis - Méditerranée, France
[2] IT Strategic Planning Group, ITRI, AIST, Japan

**Abstract.** We present a deep embedding of Bellantoni and Cook's syntactic characterization of polytime functions. We prove formally that it is correct and complete with respect to the original characterization by Cobham that required a bound to be proved manually. Compared to the paper proof by Bellantoni and Cook, we have been careful in making our proof fully contructive so that we obtain more precise bounding polynomials and more efficient translations between the two characterizations. Another difference is that we consider functions on bitstrings instead of functions on positive integers. This latter change is motivated by the application of our formalization in the context of formal security proofs in cryptography. Based on our core formalization, we have started developing a library of polytime functions that can be reused to build more complex ones.

## 1 Introduction

When formally verifying algorithms, one often proves their correctness and termination, but complexity is rarely considered. However proving correctness or termination of an algorithm that is not executable in polynomial time is of little practical use. Even at a theoretical level, it might not make much sense. For instance, in the context of security proofs one has to restrict the computational power of the adversary in the model. Indeed, an adversary with unlimited computational power could break most cryptographic schemes without actually making them insecure.

One way to take into account complexity in formal verification would be to formalize a precise execution model (e.g., Turing machines) and to explicitly count the number of steps necessary for the execution of the algorithm. Such approach would be for the least tedious and would give results depending on the particular execution model that is used whereas one is mainly interested in the complexity class independently of a particular execution model. A more convenient approach is implicit computational complexity that relates programming languages with complexity classes without relying on a particular execution model nor counting explicitly execution steps.

The main motivation behind our work presented in this paper is its application in the context of security proofs in cryptography for restricting the computational power of the adversary so that it is feasible. Cobham's thesis asserts that being feasible is the same as being computable in polynomial time [8]. Cryptographers follow Cobham's thesis in their security proofs by assuming that the adversary is computable in probabilistic polynomial time (PPT), i.e., executable on a Turing machine extended with a read-only random tape that has been filled with random bits, and working in (worst-case) polynomial time. Moreover the class of functions computable in polynomial time (a.k.a. polytime functions) has several natural closure properties that are convenient for programming. It is in particular closed under composition and a limited kind of recursion. Cobham uses those closure properties to characterize the polytime functions independently of any particular execution model. Indeed, although in his proof he uses a particular model of Turing machine, he claims that it is quite incidental, i.e., the particularities such as the number of tapes or the chosen instruction set have no significant effect on the proof. Even adding an instruction to erase a tape or put back the head in its initial position in a single step would not break the proof [8].

---

[*] Partially supported by JST-CNRS Japanese-French cooperation project on computational and symbolic proofs of security. A short version of this paper will appear in the proceedings of the 2nd International Conference on Interactive Theorem Proving (ITP 2011) [12].

[**] Supported by Kakenhi 21500046. The work presented in this paper was carried out while this author was affiliated to Research Center for Information Security, AIST.

Unfortunately, the characterization of Cobham is not fully syntactic: a size bound has to be proved on the semantics of recursive functions. This thus does not allow for an automatic procedure to check whether a program satisfies or not the conditions to be in Cobham's class. About 30 years later, Bellantoni and Cook have proposed a syntactic mechanism to control the growth rate of functions and thus eliminate the need for an explicit size bound [6]. Being a fully syntactic characterization, membership in the Bellantoni-Cook's class can be checked automatically. They show that the existence of an algorithm in Cobham's class is equivalent to the existence of an algorithm in Bellantoni-Cook's class that computes the same function. This makes their class a sound and complete characterization of polytime functions: any function definable in Bellantoni-Cook's (or Cobham's) class is computable in polynomial time, and any function computable in polynomial time is definable in Bellantoni-Cook's (and Cobham's) class.

**Related work.** It is not uncommon that a few months or a few years after a so-called security proof for a cryptographic scheme is published (e.g., in a top-level conference in cryptography), an attack on this same scheme is published. This shows that there is a need for formal verification in cryptography. This need is well-known among and acknowledged by cryptographers [11]. As a matter of fact, these last few years, several frameworks for machine-checking security proofs in cryptography have been proposed [3, 4, 16]. However, these frameworks either ignore complexity-theoretic issues or postulate the complexity of the involved functions.

Zhang has proposed a probabilistic programming language with a type system to ensure computation in probabilistic polynomial time and an equational logic to reason about those programs [25]. In [19], it has been applied to security proofs in cryptography. Zhang rely on Hofmann's SLR [13] and its extension to the probabilistic case by Mitchell et al. [15]. Those latter work are about functions on positive integers. Like us in this paper, Zhang made the move to bitstrings in order to be applicable in the context of cryptography where, for example, the bitstrings 0 and 00 are considered different although they would be identified if they were interpreted as positive integers.

In [22], it is acknowledged the need for a "polytime checker", possibly based on Bellantoni and Cook's work, and to be used to check automatically that a reduction between two NP-complete problems is computable in polynomial time. In this paper, we provide such polytime checker.

There are many other criteria to ensure that functions, defined using various programming paradigms, are in particular complexity classes. To cite only a few, some propose logical characterizations of polytime functions [14, 23] or characterizations in terms of rewrite system [2]. Others deal with different complexity classes [1]. To the best of our knowledge, none of those criteria have been applied to cryptography.

**Contributions.** In the proof assistant Coq, we have deep embedded the bitstring versions of Cobham and Bellantoni-Cook's classes and their relation. Initially, those classes were about functions on positive integers. But in the context of cryptography we must deal with bitstrings. The reformulation of Cobham's class with bitstrings and the proof that it contains exactly the function computable in polynomial time was done in [24]. In a similar way, we have reformulated the definition of Bellantoni-Cook's class.

We have also extended Bellantoni and Cook's proof that their class is equivalent to Cobham's one by making it fully constructive, i.e., we provide explicit algorithms to perform translations between the two classes. Those algorithms can be executed in Coq and extracted automatically into a certified translator in an ML dialect supported by Coq. We also make more precise the bounding polynomials thus obtaining better bounds and a more efficient translation, whereas Bellantoni and Cook overapproximate them since they are only interested in their existence and do not try to optimize the translations.

We have started to implement libraries of functions in Cobham's and Bellantoni-Cook's classes that can be used to build more complex functions.

In the context of security proofs in cryptography, we have shown how to apply our work with the second author's toolbox for certifying cryptographic primitives [16–18]. We have also extended Certicrypt [4] with support to define in Bellantoni-Cook's class the mathematical functions used by adversaries: The benefit is that one gets for free polynomials that had to be postulated before our extension, thus bringing more confidence in the security proof.

We have proved a new result on Bellantoni-Cook's class: we give explicitly a polynomial that bounds the running time of a function in Bellantoni-Cook's class. Such explicit polynomial was necessary to interface our library with Certicrypt.

**Outline.** We start by some preliminaries in Section 2. In Section 3, we formalize Cobham's and Bellantoni-Cook's classes that characterize polytime functions. Then in Sections 4 and 5 we respectively formalize the translation from Bellantoni-Cook's class to Cobham's class and vice versa. Finally in Section 6 we show how our formalization can be used for the purpose of formalizing security proofs in cryptography.

## 2 Preliminaries

In this section, we introduce our formalization of multivariate polynomials and various notations that will be used in the rest of this paper.

### 2.1 Multivariate polynomials

We have implemented a library of positive multivariate polynomials. A shallow embedding of polynomials might consist in representing them as a particular class of functions on positive integers. However, since we need in Section 4 to translate polynomials into expressions in Cobham's class, we have opted for a deep embedding. A polynomial is represented as a pair of the number of distinct variables and a list of monomials. A monomial is represented as a pair of a constant positive integer and a list of variables and their powers. A variable is represented as an integer. For example, the polynomial $3y^3 + 5x^2y + 16$ is represented by

$$(2, [(3, [(1, 3)]); (5, [(0, 2); (1, 1)]); (16, [])])$$

where the leftmost 2 is the number of variables, and variables $x$ and $y$ are respectively represented by 0 and 1. We chose to put the number of variables in the representation of a polynomial so as to easily inject a polynomial using $m$ variables into the class of polynomials with $n$ variables when $n > m$. Otherwise we would have to add artificial occurrences of variables with coefficient 0. In the library we provide utility functions in Coq to create and combine polynomials (constant, variables, addition, multiplication, composition...). We use those functions when building a polynomial. Those functions are parameterized by the number of variables, but we will omit this parameter in the rest of the paper since it will be clear from context. We write $x_0, \ldots, x_{n-1}$ for the variables of a polynomial with $n$ variables. If $P$ is a polynomial with $m$ variables and $\overline{Q} = \langle Q_0, \ldots, Q_{m-1} \rangle$ is a vector of polynomials with $n$ variables, we write $P(\overline{Q})$ for the polynomial with $n$ variables defined by substituting each variable $x_i$ in $P$ by the polynomial $Q_i$ and by applying distributivity of multiplication over addition and associativity of addition.

In [10], multivariate polynomials are represented in sparse Horner form and thus allow for a more efficient numerical evaluation of polynomials. Since we do not intend to numerically evaluate polynomials, we have opted for a more direct approach. This will moreover facilitate the connection with univariate polynomials in Certicrypt (cf. Section 6.2).

### 2.2 Notations

We list some notations that will be useful to present the results and their proofs in a concise manner. However, the meaning of those notations should be clear from the context. We write:

- $xb$ for the concatenation of the bitstring $x$ with a bit $b$ in the least significant position;
- $\overline{x}$ for a vector $\langle x_0, \ldots, x_{n-1} \rangle$ (for some $n$);
- $\overline{x}, \overline{y}$ for the concatenation of vectors $\overline{x}$ and $\overline{y}$;

- $|\overline{x}|$ for the length of a vector $\overline{x}$;
- $|x|$ for the size of a bitstring $x$;
- $\overline{|x|}$ for the vector of sizes of the components of the vector $\overline{x}$, i.e., if $\overline{x} = \langle x_0, \ldots, x_{n-1}\rangle$ then $\overline{|x|} = \langle |x_0|, \ldots, |x_{n-1}|\rangle$;
- $\overline{f(x)}$ for the vector of applications of $f$ to each component of the vector $\overline{x}$, i.e., if $\overline{x} = \langle x_0, \ldots, x_{n-1}\rangle$ then $\overline{f(x)} = \langle f(x_0), \ldots, f(x_{n-1})\rangle$;
- $\overline{f}(x)$ for the vector of applications of each component of the vector $\overline{f}$ to $x$, i.e., if $\overline{f} = \langle f_0, \ldots, f_{n-1}\rangle$ then $\overline{f}(x) = \langle f_0(x), \ldots, f_{n-1}(x)\rangle$.

## 3 Characterizing polytime functions

In this section, we explain our deep embedding of the bitstring versions of Cobham's and Bellantoni-Cook's classes, and state some of their bounding properties.

### 3.1 Cobham's class

In a seminal paper [8], Cobham characterized polytime functions as the least class of functions containing certain initial functions and closed under composition and a certain kind of recursion. However this characterization is not fully syntactic as it requires a size bound to be proved on the semantics of recursive functions.

We use the reformulation of Cobham's class taken from [24] and that deals with bitstrings instead of positive integers as it was the case in [8].

The syntax of Cobham's class $\mathcal{C}$ is given by:

$$
\begin{array}{rll}
\mathcal{C} & ::= & O \qquad\qquad\qquad \text{constant zero} \\
& | & \Pi_i^n \qquad\qquad\qquad \text{projection} \quad (i < n) \\
& | & S_b \qquad\qquad\qquad \text{successor} \\
& | & \# \qquad\qquad\qquad \text{smash} \\
& | & \mathsf{Comp}^n\ h\ \overline{g} \qquad \text{composition} \\
& | & \mathsf{Rec}\ g\ h_0\ h_1\ j \quad \text{recursion}
\end{array}
$$

where $i$ and $n$ are positive integers, $b$ is a bit, $g$, $h$, $h_0$, $h_1$ and $j$ are expressions in $\mathcal{C}$, and $\overline{g}$ is a vector of expressions in $\mathcal{C}$. Well-formed expressions $e$ in $\mathcal{C}$ have a well defined arity $\mathcal{A}(e)$ given by:

$$\mathcal{A}(O) = 0 \qquad \mathcal{A}(\Pi_i^n) = n \qquad \mathcal{A}(S_b) = 1 \qquad \mathcal{A}(\#) = 2$$

$$\frac{\mathcal{A}(h) = a_h \quad |\overline{g}| = a_h \quad \forall g \in \overline{g}, \mathcal{A}(g) = n}{\mathcal{A}(\mathsf{Comp}^n\ h\ \overline{g}) = n}$$

$$\frac{\mathcal{A}(g) = a_g \quad \mathcal{A}(h_0) = \mathcal{A}(h_1) = a_h \quad \mathcal{A}(j) = a_j \quad a_h = a_g + 2 = a_j + 1}{\mathcal{A}(\mathsf{Rec}\ g\ h_0\ h_1\ j) = a_j}$$

In our implementation, $\mathcal{A}$ is a Coq function that computes the arity of a Cobham's expression if it is well formed, or returns an error message otherwise. It is helpful when programming and debugging polytime functions in Cobham's class.

The semantics is given by:

- $O$ denotes the constant function that always returns the empty bitstring $\epsilon$.
- $\Pi_i^n(x_0, \ldots, x_{n-1})$ is equal to $x_i$.
- $S_b(x)$ is equal to $xb$.

- $\#(x, y)$ is equal to $1\underbrace{0 \ldots \ldots 0}_{|x|.|y| \text{ times}}$.
- $\mathsf{Comp}^n \; h \; \overline{g}$ is equal to the function $f$ such that:

$$f(\overline{x}) = h(\overline{g}(\overline{x}))$$

- $\mathsf{Rec} \; g \; h_0 \; h_1 \; j$ is equal to the function $f$ such that:

$$\begin{aligned} f(\epsilon, \overline{x}) &= g(\overline{x}) \\ f(yi, \overline{x}) &= h_i(y, f(y, \overline{x}), \overline{x}) \\ |f(y, \overline{x})| &\leq |j(y, \overline{x})| \end{aligned} \quad (RecBounded)$$

We illustrate Cobham's class by implementing the binary successor function:

$$\begin{array}{ll} & \mathsf{Rec} \\ \mathsf{Succ}(\epsilon) = 1 & (\mathsf{Comp}_0 \; S_1 \; O) \\ \mathsf{Succ}(x0) = x1 & (\mathsf{Comp}_0 \; S_1 \; \Pi_0^2) \\ \mathsf{Succ}(x1) = Succ(x)0 & (\mathsf{Comp}_0 \; S_0 \; \Pi_1^2) \\ \mathsf{Succ}(x) \; \leq x1 & (\mathsf{Comp}_0 \; S_1 \; \Pi_0^1) \end{array}$$

We prove in the following proposition that the output of a Cobham's function is bounded by a polynomial in the lengths of its inputs.

**Proposition 1.** *For all $f$ in $\mathcal{C}$ with a well-defined arity $\mathcal{A}(f)$ and semantics (i.e., satisfying the condition RecBounded), there exists a length-bounding monotone polynomial $\mathsf{Pol}_\mathcal{C}(f)$ such that:*

$$|f(\overline{x})| \leq (\mathsf{Pol}_\mathcal{C}(f))(\overline{|x|})$$

*Proof.* By induction on the syntax of $f$. Our proof is fully constructive in the sense that we define explicitly $\mathsf{Pol}_\mathcal{C}$. For any $f$ in $\mathcal{C}$ with arity $\mathcal{A}(f) = n$, $\mathsf{Pol}_\mathcal{C}(f)$ is the monotone polynomial with $n$ variables $x_0, \ldots, x_{n-1}$ defined by:

$$\begin{aligned} \mathsf{Pol}_\mathcal{C}(O) &= 0 \\ \mathsf{Pol}_\mathcal{C}(\Pi_i^n) &= x_i \\ \mathsf{Pol}_\mathcal{C}(S_b) &= x_0 + 1 \\ \mathsf{Pol}_\mathcal{C}(\#) &= x_0.x_1 + 1 \\ \mathsf{Pol}_\mathcal{C}(\mathsf{Comp}^n \; h \; \overline{g}) &= (\mathsf{Pol}_\mathcal{C}(h))(\overline{\mathsf{Pol}_\mathcal{C}(g)}) \\ \mathsf{Pol}_\mathcal{C}(\mathsf{Rec} \; g \; h_0 \; h_1 \; j) &= \mathsf{Pol}_\mathcal{C}(j) \end{aligned} \quad \square$$

We define a translation $\mathsf{Poly} \to \mathcal{C}$ from polynomials into Cobham's expressions. It is such that, for any polynomial $P$, $\mathsf{Poly} \to \mathcal{C}(P)$ is a unary encoding of $P$ in Cobham's class, i.e.,

$$|\mathsf{Poly} \to \mathcal{C}(P)(\overline{x})| = P(\overline{|x|})$$

### 3.2 Bellantoni-Cook's class

Bellantoni and Cook have given a fully syntactic characterization of polytime functions that does not require any explicit mechanism to count the number of computation steps [6]. The control of the growth rate of functions is achieved by distinguishing two kinds of variables: the "normal" and "safe" ones written respectively on the left and right side of a semicolon such as:

$$f(\underbrace{x_0, \ldots, x_{n-1}}_{\text{normal}}; \underbrace{x_n, \ldots, x_{n+s-1}}_{\text{safe}})$$

The syntax of Bellantoni-Cook's class $\mathcal{B}$ is given by:

$$
\begin{array}{rcll}
\mathcal{B} & ::= & 0 & \text{constant zero} \\
& | & \pi_i^{n,s} & \text{projection} \quad (i < n+s) \\
& | & s_b & \text{successor} \\
& | & \mathsf{pred} & \text{predecessor} \\
& | & \mathsf{cond} & \text{conditional} \\
& | & \mathsf{comp}^{n,s}\ h\ \overline{g_N}\ \overline{g_S} & \text{composition} \\
& | & \mathsf{rec}\ g\ h_0\ h_1 & \text{recursion}
\end{array}
$$

where $i$, $n$ and $s$ are positive integers, $b$ is a bit, $g$, $h$, $h_0$ and $h_1$ are expressions in $\mathcal{B}$, and $\overline{g_N}$ and $\overline{g_S}$ are vectors of expressions in $\mathcal{B}$. Note that, contrary to Cobham's class, a bounding function $j$ is not needed for recursion. Well-formed expressions $e$ in $\mathcal{B}$ have well defined arities $\mathcal{A}(e)$ (counting separately the numbers of normal and safe variables) given by:

$$
\mathcal{A}(0) = (0,0) \qquad \mathcal{A}(\pi_i^{n,s}) = (n,s) \qquad \mathcal{A}(s_b) = (0,1)
$$
$$
\mathcal{A}(\mathsf{pred}) = (0,1) \qquad \mathcal{A}(\mathsf{cond}) = (0,4)
$$

$$
\frac{\mathcal{A}(h) = (n_h, s_h) \quad |\overline{g_N}| = n_h \quad |\overline{g_S}| = s_h \quad \forall g \in \overline{g_N}, \mathcal{A}(g) = (n,0) \quad \forall g \in \overline{g_S}, \mathcal{A}(g) = (n,s)}{\mathcal{A}(\mathsf{comp}^{n,s}\ h\ \overline{g_N}\ \overline{g_S}) = (n,s)}
$$

$$
\frac{\mathcal{A}(g) = (n_g, s_g) \quad \mathcal{A}(h_0) = \mathcal{A}(h_1) = (n_h, s_h) \quad n_h = n_g + 1 \quad s_h = s_g + 1}{\mathcal{A}(\mathsf{rec}\ g\ h_0\ h_1) = (n_h, s_g)}
$$

This function $\mathcal{A}$ is implemented like the one for Cobham's class.

The semantics is given by:

- 0 denotes the constant function that always returns the empty bitstring $\epsilon$.
- $\pi_i^{n,s}(x_0, \ldots, x_{n-1}; x_n, \ldots, x_{n+s-1})$ is equal to $x_i$.
- $s_b(; x)$ is equal to $xb$.
- $\mathsf{pred}(; \epsilon) = \epsilon$ and $\mathsf{pred}(; xi) = x$.
- $\mathsf{cond}(; \epsilon, x, y, z) = x$, $\mathsf{cond}(; w0, x, y, z) = y$ and $\mathsf{cond}(; w1, x, y, z) = z$.
- $\mathsf{comp}^{n,s}\ h\ \overline{g_N}\ \overline{g_S}$ is equal to the function $f$ such that:

$$f(\overline{x}; \overline{y}) = h(\overline{g_N}(\overline{x};); \overline{g_S}(\overline{x}; \overline{y}))$$

Note here that the functions in $\overline{g_N}$ only have access to normal variables.
- $\mathsf{rec}\ g\ h_0\ h_1$ is equal to the function $f$ such that:

$$
\begin{array}{rl}
f(\epsilon, \overline{x}; \overline{y}) &= g(\overline{x}; \overline{y}) \\
f(zi, \overline{x}; \overline{y}) &= h_i(z, \overline{x}; f(z, \overline{x}; \overline{y}), \overline{y})
\end{array}
$$

Note here that the result of the recursive call $f(z, \overline{x}; \overline{y})$ is passed at a safe position. This prevents it to be used as the recursion argument in a nested recursion.

One can see that, contrary to Cobham's class $\mathcal{C}$, there is no size bound to be proved on recursive functions: Bellantoni-Cook's class $\mathcal{B}$ is syntactically defined.

Reader may have noticed that our definition of Bellantoni-Cook's class is slightly different from the one in [6]. First, here the conditional $\mathsf{cond}$ distinguishes between three cases (empty, even or odd bitstrings), whereas in [6] the empty bitstring is treated as an even one. Second, here the base case for recursion is the empty bitstring, whereas in [6] it is any bitstring whose interpretation as a positive integer is 0, i.e., the empty bitstring or any bitstring made of any number of bits 0 only. We made those changes because in cryptography one wants to distinguish, for example, bitstrings 0 and 00 although they would have the same interpretation in terms of positive integers. These changes are validated by the results we proved in the rest of the paper where we translate our Bellantoni-Cook's expressions to/from the bitstring version of Cobham's expressions [24].

The following examples illustrate how one can program addition and multiplication in Bellantoni-Cook's class and their respective arity:

$$
\begin{array}{ll}
plus := \text{rec} & mult := \text{rec} \\
\quad (\pi_0^{0,1}) & \quad (\text{comp}^{1,0}\ O\ \langle\rangle\ \langle\rangle) \\
\quad (\text{comp}^{1,2}\ S_1\ \langle\rangle\ \langle\pi_1^{1,2}\rangle) & \quad (\text{comp}^{1,2}\ plus\ \langle\pi_1^{2,0}\rangle\ \langle\pi_2^{2,1}\rangle) \\
\quad (\text{comp}^{1,2}\ S_1\ \langle\rangle\ \langle\pi_1^{1,2}\rangle) & \quad (\text{comp}^{1,2}\ plus\ \langle\pi_1^{2,0}\rangle\ \langle\pi_2^{2,1}\rangle)
\end{array}
$$

$$\mathcal{A}(plus) = (1,1) \qquad \mathcal{A}(mult) = (2,0)$$

We prove in the following proposition that the output of a Bellantoni-Cook's function is bounded by the sum of a polynomial in the lengths of its normal inputs and the size of its longest safe input. This is so because syntactic restrictions ensure that we cannot increase the lengths of safe inputs by more than an additive constant that will be taken into account in the polynomial part.

**Proposition 2 (Polymax Bounding).** *For all $f$ in $\mathcal{B}$ with well-defined arities $\mathcal{A}(f)$, there exists a length-bounding monotone polynomial $\mathsf{Pol}_\mathcal{B}(f)$ such that, for all $\overline{x}$ and $\overline{y}$:*

$$|f(\overline{x};\overline{y})| \leq (\mathsf{Pol}_\mathcal{B}(f))(\overline{|x|}) + \max_i |y_i|$$

*Proof.* By induction on the syntax of $f$. Our proof is fully constructive in the sense that we define explicitly $\mathsf{Pol}_\mathcal{B}$. For any $f$ in $\mathcal{B}$ with arity $\mathcal{A}(f) = (n,s)$, $\mathsf{Pol}_\mathcal{B}(f)$ is the monotone polynomial with $n$ variables $x_0, \ldots, x_{n-1}$ defined by:

$$
\begin{array}{lll}
\mathsf{Pol}_\mathcal{B}(0) & = & 0 \\
\mathsf{Pol}_\mathcal{B}(\pi_i^{n,s}) & = & x_i \text{ if } i < n \\
& & 0 \text{ otherwise} \\
\mathsf{Pol}_\mathcal{B}(s_b) & = & 1 \\
\mathsf{Pol}_\mathcal{B}(\mathsf{pred}) & = & 0 \\
\mathsf{Pol}_\mathcal{B}(\mathsf{cond}) & = & 0 \\
\mathsf{Pol}_\mathcal{B}(\mathsf{comp}^{n,s}\ h\ \overline{g_N}\ \overline{g_S}) & = & \mathsf{Pol}_\mathcal{B}(h)(\overline{\mathsf{Pol}_\mathcal{B}(g_N)}) + \sum(\overline{\mathsf{Pol}_\mathcal{B}(g_S)}) \\
\mathsf{Pol}_\mathcal{B}(\mathsf{rec}\ g\ h_0\ h_1) & = & \mathsf{shift}(\mathsf{Pol}_\mathcal{B}(g)) + x_0.(\mathsf{Pol}_\mathcal{B}(h_0) + \mathsf{Pol}_\mathcal{B}(h_1))
\end{array}
$$

where $\mathsf{shift}(P)$ is the polynomial $P$ with each variable $x_i$ replaced by $x_{i+1}$. □

We define a translation $\mathsf{Poly} \to \mathcal{B}$ from polynomials into Bellantoni-Cook's expressions. It is such that, for any polynomial $P$, $\mathsf{Poly} \to \mathcal{B}(P)$ is a unary encoding of $P$ in Bellantoni-Cook's class, i.e.,

$$|\mathsf{Poly} \to \mathcal{B}(P)(\overline{x})| = P(\overline{|x|})$$

In order to ease further development of functions in Bellantoni-Cook's class, we provide a mechanism to infer automatically the optimal values for the parameters $n$ and $s$ appearing in $\pi_i^{n,s}$ and $\mathsf{comp}^{n,s}$ thus obtaining more elegant code. This is implemented with a new syntax $\mathcal{B}_{\mathsf{inf}}$ for Bellantoni-Cook's class where arities do not appear in the syntax:

$$
\begin{array}{lll}
\mathcal{B}_{\mathsf{inf}} ::= & 0 & \text{constant zero} \\
& |\ \pi_i^{\mathsf{normal}} & \text{projection (normal)} \quad (i < n) \\
& |\ \pi_i^{\mathsf{safe}} & \text{projection (safe)} \qquad (i < s) \\
& |\ s_b & \text{successor} \\
& |\ \mathsf{pred} & \text{predecessor} \\
& |\ \mathsf{cond} & \text{conditional} \\
& |\ \mathsf{comp}\ h\ \overline{g_N}\ \overline{g_S} & \text{composition} \\
& |\ \mathsf{rec}\ g\ h_0\ h_1 & \text{recursion}
\end{array}
$$

We have validated this new syntax by providing certified translations between $\mathcal{B}_{\mathsf{inf}}$ and $\mathcal{B}$. When translating from $\mathcal{B}_{\mathsf{inf}}$ to $\mathcal{B}$, we can force arities to be larger that the minimal ones that are inferred.

## 4 Compiling Bellantoni-Cook into Cobham

In this section, we provide our formalization of the translation of expressions in Bellantoni-Cook's class into expressions in Cobham's class. The main result is stated in the following theorem.

**Theorem 1.** *For all $f$ in $\mathcal{B}$ with well defined arities $\mathcal{A}(f)$, there exists $f'$ in $\mathcal{C}$ such that for all vectors of bitstrings $\overline{x}$ and $\overline{y}$, $f(\overline{x}; \overline{y}) = f'(\overline{x}, \overline{y})$.*

*Proof.* The proof is split into two inductions on the syntax of $f$: The first one to prove the equality and the second one to prove that the condition *RecBounded* is satisfied. Our proof is fully constructive in the sense that we define explicitly the translation $\mathcal{B} \to \mathcal{C}$ from $\mathcal{B}$ to $\mathcal{C}$ and define $f'$ as $\mathcal{B}{\to}\mathcal{C}(f)$. The difficulty of the proof is in the generation of a Cobham expression that satisfies the condition *RecBounded*, and to build the polynomial $j$ that bounds the recursive calls.

- The first cases are immediate:
$$\begin{array}{rcl}
\mathcal{B}{\to}\mathcal{C}(0) & = & O \\
\mathcal{B}{\to}\mathcal{C}(\pi_i^{n,s}) & = & \Pi_i^{n+s} \\
\mathcal{B}{\to}\mathcal{C}(s_b) & = & S_b
\end{array}$$

- pred and cond are translated by using Rec:
$$\begin{array}{rcl}
\mathcal{B}{\to}\mathcal{C}(\mathsf{pred}) & = & \mathsf{Rec}\ O\ \Pi_0^2\ \Pi_0^2\ \Pi_0^1 \\
\mathcal{B}{\to}\mathcal{C}(\mathsf{cond}) & = & \mathsf{Rec}\ \Pi_0^3\ \Pi_4^5\ \Pi_3^5 \\
& & \mathsf{Comp}^4\ \#\ \langle \\
& & \quad \mathsf{Comp}^4\ S_1\ \langle \Pi_1^4 \rangle; \\
& & \quad \mathsf{Comp}^4\ \#\ \langle\ \mathsf{Comp}^4\ S_1\ \langle \Pi_2^4 \rangle;\ \mathsf{Comp}^4\ S_1\ \langle \Pi_3^4 \rangle\ \rangle\rangle
\end{array}$$

- For $\mathsf{comp}^{n,s}\ h\ \overline{g_N}\ \overline{g_S}$ we need to add dummy variables, since the functions in $\overline{g_N}$ do not take the safe arguments as parameters. We need to transform these functions in $\overline{g_N}$ into functions with arity $n + s$. $\mathsf{dummies}_s$ (written in $\mathcal{C}$) add $s$ dummy variables that are ignored:
$$\mathcal{B}{\to}\mathcal{C}(\mathsf{comp}^{n,s}\ h\ \overline{g_N}\ \overline{g_S}) = \mathsf{Comp}^{n+s}\ \dfrac{\mathcal{B}{\to}\mathcal{C}(h)}{\left(\overline{\mathsf{dummies}_s\ (\mathcal{B}{\to}\mathcal{C}(g_N))}, \overline{\mathcal{B}{\to}\mathcal{C}(g_S)}\right)}$$

- For $\mathsf{rec}\ g\ h_0\ h_1$ we need to change the order of the arguments passed to the translations of $h_0$ and $h_1$. Indeed, while in $\mathcal{B}$ the recursive argument is put after the normal ones, in $\mathcal{C}$ it should be the second argument. This reordering is done by the function $\mathsf{move\_arg}_{2,n}$ (written in $\mathcal{C}$). Moreover, we need to derive a suitable bound for the fourth argument of Rec. By Proposition 2 and the fact that the sum $|x_n| + \cdots + |x_{n+s-1}|$ of the sizes of the safe arguments is greater than or equal to the maximum size of the safe arguments, we can take the polynomial $\mathsf{Pol}_{\mathcal{B}}(\mathsf{rec}\ g\ h_0\ h_1) + x_n + \cdots + x_{n+s-1}$ for the bound. We then use $\mathsf{Poly}{\to}\mathcal{C}$ to encode it in $\mathcal{C}$:

$$\begin{array}{rl}
\mathcal{B}{\to}\mathcal{C}(\mathsf{rec}\ g\ h_0\ h_1) = & \mathsf{Rec} \\
& \quad \mathcal{B}{\to}\mathcal{C}(g) \\
& \quad \mathsf{move\_arg}_{2,n}\ (\mathcal{B}{\to}\mathcal{C}(h_0)) \\
& \quad \mathsf{move\_arg}_{2,n}\ (\mathcal{B}{\to}\mathcal{C}(h_1)) \\
& \quad \mathsf{Poly}{\to}\mathcal{C} \begin{pmatrix} \mathsf{Pol}_{\mathcal{B}}(\mathsf{rec}\ g\ h_0\ h_1)\ + \\ x_n\ +\ \cdots\ +\ x_{n+s-1} \end{pmatrix} \quad \square
\end{array}$$

## 5 Compiling Cobham into Bellantoni-Cook

Contrary to Bellantoni-Cook's class $\mathcal{B}$, one does not distinguish between normal and safe variables in Cobham's class $\mathcal{C}$. In $\mathcal{C}$ it is possible to recur on any argument, whereas in $\mathcal{B}$ one can only recur on normal arguments. Thus, when translating from $\mathcal{C}$ to $\mathcal{B}$, we must introduce a distinction and deal appropriately with recursion. In our formalization, we follow Bellantoni and Cook's translation scheme by assuming that all the arguments are safe and adding an artificial normal argument $w$ whose length is large enough to ensure enough recursion steps. This gives us the lemma below. After that, we will get rid of $w$ thus obtaining Theorem 2.

**Lemma 1 (Recursion Simulation).** *For all function $f$ in $\mathcal{C}$ with well-defined arity $\mathcal{A}(f) = n$ and semantics (i.e., satisfying the condition RecBounded), there exists an $f'$ in $\mathcal{B}$ and a monotone polynomial $\mathsf{Pol}_{\mathcal{C} \to \mathcal{B}}(f)$ such that for all vector of bitstrings $\overline{x}$ and bitstring $w$ such that $\mathsf{Pol}_{\mathcal{C} \to \mathcal{B}}(f)(\overline{|x|}) \leq |w|$, $f(\overline{x}) = f'(w; \overline{x})$.*

*Proof.* By induction on the syntax of $f$. Our proof is fully constructive in the sense that we define explicitly the polynomial $\mathsf{Pol}_{\mathcal{C} \to \mathcal{B}}(f)$ and the translation $\mathcal{C} \to \mathcal{B}$, and define $f'$ as $\mathcal{C} \to \mathcal{B}(f)$. Our translation is such that if $\mathcal{A}(f) = n$ then $\mathcal{A}(f') = (1, n)$, i.e., $f'$ takes one normal argument and $n$ safe arguments.

- The first cases for $\mathcal{C} \to \mathcal{B}(f)$ are immediate. We just have to make sure that the arities are right:

$$\begin{aligned}
\mathcal{C} \to \mathcal{B}(O) &= \mathsf{comp}^{1,n}\ O\ \langle\rangle\ \langle\rangle \\
\mathcal{C} \to \mathcal{B}(\Pi_i^n) &= \pi_{i+1}^{1,n} \\
\mathcal{C} \to \mathcal{B}(S_b) &= \mathsf{comp}^{1,n}\ s_b\ \langle\rangle\ \langle \pi_1^{1,n} \rangle \\
\mathcal{C} \to \mathcal{B}(\mathsf{Comp}^n\ h\ \overline{g}) &= \mathsf{comp}^{1,n}\ (\mathcal{C} \to \mathcal{B}(h))\ \langle \pi_1^{1,0} \rangle\ \overline{\mathcal{C} \to \mathcal{B}(g)}
\end{aligned}$$

$\mathsf{Pol}_{\mathcal{C} \to \mathcal{B}}(f)(\overline{|x|})$ is also immediate for these first cases:

$$\begin{aligned}
\mathsf{Pol}_{\mathcal{C} \to \mathcal{B}}(O) &= 0 \\
\mathsf{Pol}_{\mathcal{C} \to \mathcal{B}}(\Pi_i^n) &= 0 \\
\mathsf{Pol}_{\mathcal{C} \to \mathcal{B}}(S_b) &= 0 \\
\mathsf{Pol}_{\mathcal{C} \to \mathcal{B}}(\mathsf{Comp}^n\ h\ \overline{g}) &= \mathsf{Pol}_{\mathcal{C} \to \mathcal{B}}(h)(\overline{\mathsf{Pol}_{\mathcal{C}}(g)}) + \sum_{g \in \overline{g}} \mathsf{Pol}_{\mathcal{C} \to \mathcal{B}}(g)
\end{aligned}$$

The rightness of the case for $\mathsf{Comp}$ follows by induction hypothesis and Proposition 1.

- For the case of $\mathsf{Rec}\ g\ h_0\ h_1\ j$, we follow [6] in defining intermediate functions in $\mathcal{B}$. However we need less of them since our definition of $\hat{f}$ below is simpler. We define $P$ in $\mathcal{B}$ such that $P(a; b)$ removes the $|a|$ least significant bits of $b$, i.e., $P(\epsilon; b) = b$ and $P(ai; b) = \mathsf{pred}(; P(a; b))$. We define $Y$ in $\mathcal{B}$ such that $Y(z, w; y)$ removes the $|w| - |z|$ least significant bits of $y$, i.e., $Y(z, w; y) = P(P'(z, w); y)$ where $P'(a, b;) = P(a; b)$. $P$ and $Y$ are then used to define $\hat{f}$ in $\mathcal{B}$:

$$\hat{f}(\epsilon, w; y, \overline{x}) = g(w; \overline{x})$$
$$\hat{f}(zj, w; y, \overline{x}) = \begin{cases} g'(w; \overline{x}) & \text{if } Y(S_1 z, w; y) \text{ is } \epsilon \\ h_0'(w, Y(z, w; y), \hat{f}(z, w; y, \overline{x}), \overline{x}) & \text{if } Y(S_1 z, w; y) \text{ is even} \\ h_1'(w; Y(z, w; y), \hat{f}(z, w; y, \overline{x}), \overline{x}) & \text{if } Y(S_1 z, w; y) \text{ is odd} \end{cases}$$

where $g'$, $h_0'$ and $h_1'$ are respectively $\mathcal{C} \to \mathcal{B}(g)$, $\mathcal{C} \to \mathcal{B}(h_0)$ and $\mathcal{C} \to \mathcal{B}(h_1)$. Our definition of $\hat{f}$ is simpler than in [6] because we do not need, like in [6], an additional intermediate function to check whether $y$ is an encoding of the positive integer 0. In our case, we stop the recursion when $y$ is equal to $\epsilon$, since the $\mathsf{cond}$ can check whether the first safe argument is $\epsilon$.

We then define $f'(w; y, \overline{x})$ in $\mathcal{B}$ such that it is equal to $\hat{f}(w, w; y, \overline{x})$, and finally:

$$\mathcal{C} \to \mathcal{B}(\mathsf{Rec}\ g\ h_0\ h_1\ j) = f'\ (\mathcal{C} \to \mathcal{B}(g))\ (\mathcal{C} \to \mathcal{B}(h_0))\ (\mathcal{C} \to \mathcal{B}(h_1))$$

For the polynomial, we have:

$$\begin{aligned}
\mathsf{Pol}_{\mathcal{C} \to \mathcal{B}}(\mathsf{Rec}\ g\ h_0\ h_1\ j)(|y|, \overline{|x|}) = &(\mathsf{Pol}_{\mathcal{C} \to \mathcal{B}}(h_0) + \mathsf{Pol}_{\mathcal{C} \to \mathcal{B}}(h_1))(|y|, \mathsf{Pol}_{\mathcal{C}}(f), \overline{|x|}) + \\
&\mathsf{shift}(\mathsf{Pol}_{\mathcal{C} \to \mathcal{B}}(g)(\overline{|x|})) + |y| + 2
\end{aligned}$$

- We can define the smash function # from $\mathcal{C}$ in $\mathcal{B}$ by a double recursion:

$$\begin{aligned}
\#'(\epsilon, y) &= y \\
\#'(xi, y) &= \#'(x, y)\, 0 \quad \text{(concatenation with a bit 0)} \\
\#(\epsilon, y) &= 1 \\
\#(xi, y) &= \#'(y, \#(x, y))
\end{aligned}$$

In order to implement in $\mathcal{B}$ those two recursive functions we apply the same technique as in the case of Rec above. We first obtain a $\#'$ in $\mathcal{B}$ by constructing $f'$ with $g = \pi_1^{1,1}$ and $h_0 = h_1 = \mathsf{comp}^{1,3}\ S_0\ \langle\rangle\ \langle\pi_2^{1,3}\rangle$. We then obtain $\#$ in $\mathcal{B}$ by applying the same construction of $f'$ with $g = \mathsf{one}^{1,1}$ and $h_0 = h_1 = \mathsf{dummies}_{0,1}(\mathsf{comp}^{1,2}\ \#'\ \langle\pi_0^{1,0}\rangle\ \langle\pi_2^{1,2}; \pi_1^{1,2}\rangle)$.

For the polynomial, we obtain (after simplification):

$$\mathsf{Pol}_{\mathcal{C}\to\mathcal{B}}(\#) = x_0 + 2x_1 + 18 \qquad \square$$

Finally, the main result of this section is stated in the following theorem.

**Theorem 2.** *For all $f$ in $\mathcal{C}$ with well-defined arity $\mathcal{A}(f)$ and semantics (i.e., satisfying the condition RecBounded), there exists $f'$ in $\mathcal{B}$ such that, for all vectors of bitstrings $\overline{x}$, $f(\overline{x}) = f'(\overline{x}; )$.*

*Proof.* We define the expression $b_f$ in $\mathcal{B}$ by:

$$\mathsf{Poly}\to\mathcal{B}(\mathsf{Pol}_{\mathcal{C}\to\mathcal{B}}(f))$$

By definition of $\mathsf{Poly}\to\mathcal{B}$, for all vector of bistrings $\overline{x}$, $|b_f(\overline{x}; )| = \mathsf{Pol}_{\mathcal{C}\to\mathcal{B}}(f)(\overline{|x|})$. We can thus apply Lemma 1 which gives:

$$f'(\overline{x};) = (\mathcal{C}\to\mathcal{B}(f))(b_f(\overline{x};); \overline{x}) \qquad \square$$

Our translation gives a more efficient code than the one in [6] since our definition of $b_f$ is more precise: the number of recursive calls will be no more than what is strictly necessary. Indeed, authors of [6] use general properties of multivariate polynomials to first prove the existence of positive integers $a$ and $c$ such that

$$\mathsf{Pol}_{\mathcal{C}\to\mathcal{B}}(f)(\overline{|x|}) \leq \left(\sum_j |x_j|\right)^a + c$$

and then use $a$ and $c$ to build a $b_f$ that satisfies the condition of Lemma 1, i.e., $\mathsf{Pol}_{\mathcal{C}\to\mathcal{B}}(f)(\overline{|x|}) \leq |b(\overline{x};)|$. Their $b_f$ is an overapproximation of $\mathsf{Pol}_{\mathcal{C}\to\mathcal{B}}(f)$ while our $b_f$ is an exact encoding.

## 6 Applications

Security properties in cryptography are often modeled as games, and then security proofs consist in showing that no adversary can win the game [7, 21]. Most of those proofs are based on computational assumptions that state that an effective adversary cannot solve a particular mathematical problem, e.g., Diffie-Hellman problems [9]. Effective adversaries are modeled as strict probabilistic polynomial-time functions, i.e., independently of the random choices, the execution time is bounded by a polynomial in a security parameter (typically the length of the inputs). This means that an adversary can be modeled as a polytime function with, as an additional parameter, a long enough bitstring that will be used by the adversary as its source of random bits.

### 6.1 Application to the second author's toolbox

The second author's toolbox is a collection of definitions and lemmas to be used for verifying game transformations in security proofs [16–18]. With this toolbox, our library can be used as such when applying a computational hypothesis. The computational hypotheses can indeed be restricted to adversaries defined in Cobham's or Bellantoni-Cook's class (it is not too restrictive since those classes are complete) and adversaries appearing in proofs must then be defined in one of those classes. For example, when applying the Decisional Diffie-Hellman assumption (DDH) in the security proof for Hashed ElGamal in [16], a new adversary $\varphi$ is built from two adversaries $A_1$ and $A_2$:

$$\varphi(X, Y, Z) =_{def} A_2(r, (X, k), (Y, H_k(Z) \oplus \pi_b(A_1(r, (X, k))))) \stackrel{?}{=} b$$

where $b$, $k$ and $r$ are fixed, $\oplus$ is the bitwise exclusive or (xor), $\pi_b$ is the $b^{th}$ projection ($b$ is equal to 1 or 2), and $\stackrel{?}{=}$ is the equality test. That $A_1$ and $A_2$ are polytime is given by hypothesis. We can also assume that the hash function $H_k$ is polytime. Being polytime, they are definable in Bellantoni-Cook's class. Moreover, projections, exclusive or and equality test are easily defined in Bellantoni-Cook's class. Therefore $\varphi$ is easily definable in Bellantoni-Cook's class and thus it is polytime.

### 6.2 Application to Certicrypt

The application of our library to Certicrypt requires more work but brings noticeable benefits.

In Certicrypt, a game is a probabilistic imperative program that transforms a distribution of input states into a distribution of output states. A state includes a time index. A distribution of states is polynomially bounded if there are two (univariate) polynomials $p$ and $q$ respectively bounding the size of the data and the time index of each state in the distribution. A program is strict probabilistic polynomial time (PPT) iff: it always terminates; and, there exists two (univariate) polynomial transformers $F$ and $G$ such that, for every polynomially bounded (by $p$ and $q$) distribution of input states, the distribution of output states is bounded by $F(p)$ (bounding the output size) and $q + G(p)$ (bounding the execution time). Interested reader should refer to [4] for further explanation about this way to formalize PPT.

We have built an interface with Certicrypt made of the following components:

- The core language of Certicrypt can be extended with user-defined types and functions. But the time cost of each function has to be axiomatized in the current implementation of Certicrypt. We have added the possibility to include functions that have been defined in our implementation of Bellantoni-Cook's class and that are thus automatically proved executable in polynomial time, thus removing the need for postulates.
- We have added a conversion from any multivariate polynomial $P$ given by our library into a univariate one $\lceil P \rceil$ in Certicrypt that overapproximates $P$ when applied to the maximal argument: This is easily done by substituting all variables $x_0$, ..., $x_{n-1}$ in $P$ by a single variable $x$:

$$\lceil P \rceil =_{def} P[x_0 \mapsto x; \ldots; x_{n-1} \mapsto x]$$

- In the case of a program $c$ defined in Bellantoni-Cook's class, we can take $F(p)$ to be equal to:

$$1 + 2 \lceil \mathsf{Pol}_\mathcal{B}(c) \rceil (p)$$

  This is justified by Proposition 2. The multiplication by 2 and and addition of 1 are here because of technical reasons coming from Certicrypt. For example, the multiplication by 2 comes from the fact that the size of a boolean in Certicrypt is 2.
- In order to obtain $G(p)$, we need to consider the obvious implementation of Bellantoni-Cook's class on a stack machine as described in Section 3.4.2 of [5]. We have equipped the semantics of Bellantoni-Cook's class with a time index that keeps track of the running time. We can then prove that the multivariate polynomial $\mathsf{Pol}_{\mathsf{time}}$ below is an upper bound of the running time, and use it to define $G(p)$.

$$\begin{aligned}
\mathsf{Pol}_{\mathsf{time}}(0) = \mathsf{Pol}_{\mathsf{time}}(\pi_i^{n,s}) &= \mathsf{Pol}_{\mathsf{time}}(s_b) = \mathsf{Pol}_{\mathsf{time}}(\mathsf{pred}) = \mathsf{Pol}_{\mathsf{time}}(\mathsf{cond}) = 1 \\
\mathsf{Pol}_{\mathsf{time}}(\mathsf{comp}^{n,s}\ h\ \overline{g_N}\ \overline{g_S}) &= \mathsf{Pol}_{\mathsf{time}}(h)(\overline{\mathsf{Pol}_{\mathcal{B}}(g_N)}) + \\
&\quad \sum(\overline{\mathsf{Pol}_{\mathsf{time}}(g_N)}) + \sum(\overline{\mathsf{Pol}_{\mathsf{time}}(g_S)}) \\
\mathsf{Pol}_{\mathsf{time}}(\mathsf{rec}\ g\ h_0\ h_1) &= \mathsf{shift}(\mathsf{Pol}_{\mathsf{time}}(g)) + \\
&\quad x_0.(\mathsf{Pol}_{\mathsf{time}}(h_0) + \mathsf{Pol}_{\mathsf{time}}(h_1))
\end{aligned}$$

where $\mathsf{shift}(P)$ is the polynomial $P$ with each variable $x_i$ replaced by $x_{i+1}$. An interesting thing about $\mathsf{Pol}_{\mathsf{time}}$ is that, in the case of $\mathsf{comp}^{n,s}\ h\ \overline{g_N}\ \overline{g_S}$, it is necessary to consider the size $\overline{\mathsf{Pol}_{\mathcal{B}}(g_N)}$ of the outputs of the functions in $\overline{g_N}$ for the running time of $h$, but not the size of the outputs of the functions in $\overline{g_S}$. This is so because syntactic restrictions ensure that we cannot increase the lengths of safe inputs by more than an additive constant. Finally, for a program $c$ defined in Bellantoni-Cook's class, we take $G(p)$ to be equal to:

$$\lceil \mathsf{Pol}_{\mathsf{time}}(c) \rceil (p)$$

The reader might be surprised that in this section we consider a particular implementation of Bellantoni-Cook's class while in introduction we said that we are interested in complexity independently of any execution model. The reason is that, although being in Cobham's or Bellantoni-Cook's class guarantees that there exists a polynomial bounding the execution time, it does not give any clue on the actual value of this polynomial. However Certicrypt requires that we explicitly give such a polynomial. This is why we consider here a particular execution model: to be able to compute a polynomial.

## 7 Conclusions and future work

We have formalized Cobham's and Bellantoni-Cook's classes and their relations in the proof assistant Coq. Usage of proof assistant led us to formalize parts of the proofs that were only informal in Bellantoni and Cook's paper. Our formalization allows to use those classes as programming languages to define any function that is computable in polynomial time. We have shown in particular that it can be used to build adversaries in security proofs in cryptography.

**Future Work.** In order to facilitate the use of our formalization, an important future work is to carry on developing a convenient library of polytime functions on bitstrings that can be reused in the construction of more advanced polytime functions. It is easy to implement bitwise operations such as bitwise XOR, NOT, AND, etc. However, when implementing numerical operations such as bitwise addition, dealing with the carry bit does not fit immediately in Bellantoni-Cook's recursion scheme. One possible solution is to implement binary addition, multiplication and other numerical functions in Cobham's class such as in [20], and use our automatic translation $\mathcal{C} \to \mathcal{B}$ (defined in Section 5) to derive their implementations in Bellantoni-Cook's class. Also, our approach can be extended with higher order so as to formalize a more powerful programming language such as CSLR [25].

**Acknowledgments.** We are grateful to Benjamin Grégoire and Santiago Zanella Béguelin for replying so quickly to every question we have had on Certicrypt, and for their comments on an earlier draft of this paper. We thank anonymous reviewers for their helpful suggestions, all of which improved the paper.

## References

1. T. Arai, N. Eguchi. A new function algebra of EXPTIME functions by safe nested recursion. ACM Transactions on Computational Logic 10(4). ACM, 2009.

2. M. Avanzini, G. Moser. Complexity Analysis by Rewriting. In *Proceedings of the 9th International Symposium on Functional and Logic Programming (FLOPS 2008)*, volume 4989 of Lecture Notes in Computer Science, pages 130–146. Springer, 2008.
3. M. Backes, M. Berg, and D. Unruh. A formal language for cryptographic pseudocode. In *Proceedings of the 15th International Conference on Logic for Programming, Artificial Intelligence and Reasoning (LPAR 2008)*, volume 5330 of Lecture Notes in Computer Science, pages 353–376. Springer, 2008.
4. G. Barthe, B. Grégoire, and S. Zanella Béguelin. Formal certification of code-based cryptographic proofs. In *Proceedings of the 36th ACM SIGPLAN- SIGACT Symposium on Principles of Programming Languages (POPL 2009)*, pages 90–101. ACM, 2009.
5. S. Bellantoni. *Predicative Recursion and Computational Complexity*. PhD Thesis, University of Toronto, 1992.
6. S. Bellantoni and S. A. Cook. A new recursion-theoretic characterization of the polytime functions. In *Computational Complexity*, 2:97–110, 1992.
7. M. Bellare and P. Rogaway. The security of triple encryption and a framework for code-based game-playing proofs. In *Advances in Cryptology - EUROCRYPT06*, volume 4004 of Lecture Notes in Computer Science, pages 409–426. Springer, 2006.
8. A. Cobham. The intrinsic computational difficulty of functions. In *Proceedings of the 1964 International Congress for Logic, Methodology, and the Philosophy of Science*, pages 24–30. North Holland, 1964.
9. W. Diffie and M. E. Hellman. New directions in cryptography. In *IEEE Transactions on Information Theory*, IT-22(6):644–654. IEEE Computer Society, 1976.
10. B. Grégoire and A. Mahboubi. Proving Equalities in a Commutative Ring Done Right in Coq. In *Proceedings of the 18th International Conference on Theorem Proving in Higher Order Logics (TPHOLs 2005)*, volume 3603 of Lecture Notes in Computer Science, pages 98–113. Springer, 2005.
11. S. Halevi. A plausible approach to computer-aided cryptographic proofs. Cryptology ePrint Archive, Report 2005/181, 2005.
12. S. Heraud and D. Nowak. A formalization of polytime functions. In*Proceedings of the 2nd International Conference on Interactive Theorem Proving (ITP 2011)*, Lecture Notes in Computer Science. Springer, 2011. To appear.
13. M. Hofmann. Safe recursion with higher types and BCK-algebra. In *Annals of Pure and Applied Logic*, 104(1-3):113–166, 2000.
14. D. Leivant. A foundational delineation of computational feasibility. In *Sixth Annual IEEE Symposium on Logic in Computer Science*, pages 2–11. IEEE Computer Society, 1991
15. J. C. Mitchell, M. Mitchell, and A. Scedrov. A linguistic characterization of bounded oracle computation and probabilistic polynomial time. In *Proceedings of the 39th Annual Symposium on Foundations of Computer Science (FOCS'98)*, pages 725–733. IEEE Computer Society, 1998.
16. D. Nowak. A framework for game-based security proofs. In *Proceedings of the 9th International Conference on Information and Communications Security (ICICS 2007)*, volume 4861 of Lecture Notes in Computer Science, pages 319–333. Springer, 2007.
17. D. Nowak. On formal verification of arithmetic-based cryptographic primitives. In *Proceedings of the 11th International Conference on Information Security and Cryptology (ICISC 2008)*, volume 5461 of Lecture Notes in Computer Science, pages 368–382. Springer, 2008.
18. D. Nowak. A Certification Toolbox for Cryptographic Algorithms, Coq development, http://staff.aist.go.jp/david.nowak/toolbox/, last access: May 31, 2011.
19. D. Nowak and Y. Zhang. A calculus for game-based security proofs. In *Proceedings of the 4th International Conference on Provable Security (ProvSec 2010)*, volume 6402 of Lecture Notes in Computer Science, pages 35-52. Springer, 2010.
20. H. E. Rose. *Subrecursion: functions and hierarchies*, Oxford Logic Guides 9, Clarendon Press, Oxford, 1984.
21. V. Shoup. Sequences of games: A tool for taming complexity in security proofs. Cryptology ePrint Archive, Report 2004/332, 2004.
22. C. Schürmann and J. Shah. Representing reductions of NP-complete problems in logical frameworks: A case study. In *Proceedings of the Eighth ACM SIGPLAN International Conference on Functional Programming, Workshop on Mechanized reasoning about languages with variable binding (MERLIN 2003)*. ACM, 2003.
23. C. Schürmann and J. Shah. Identifying Polynomial-Time Recursive Functions. In *Proceedings of the 19th International Workshop,14th Annual Conference of the EACSL (CSL 2005)*, volume 3634 of Lecture Notes in Computer Science, pages 525–540. Springer, 2005.
24. G. J. Tourlakis. *Computability*, Reston, 1984.
25. Y. Zhang. The computational SLR: a logic for reasoning about computational indistinguishability. In *Mathematical Structures in Computer Science*, 20: 951-975. Cambridge University Press, 2010.